\documentclass[sn-mathphys,refee]{sn-jnl-no}

\jyear{2021}%

\theoremstyle{thmstyleone}%
\theoremstyle{thmstyletwo}%
\usepackage{amsmath}
\usepackage{amssymb}
\usepackage{graphics}
\usepackage{graphicx}
\usepackage{epstopdf}
\usepackage{dcolumn}
\usepackage{bm}
\usepackage{xcolor}
\usepackage{multirow}
\usepackage{caption}
\usepackage{setspace}

\theoremstyle{thmstylethree}%
\begin{document}

\title[Article]{\textbf{Incoherent mode division multiplexing for high-security information encryption}}

\author[1,2]{\fnm{Xin} \sur{Liu}}

\author*[3,4]{\fnm{Sergey A.} \sur{Ponomarenko}}
\email{serpo@dal.ca}

\author[5]{\fnm{Fei} \sur{Wang}}

\author*[1,2]{\fnm{Yangjian} \sur{Cai}}
\email{yangjian\_cai@163.com}

\author*[1,2]{\fnm{Chunhao} \sur{Liang}}
\email{chunhaoliang@sdnu.edu.cn}

\affil[1]{Shandong Provincial Engineering and Technical Center of Light Manipulation \& Shandong Provincial Key Laboratory of Optics and Photonic Devices, School of Physics and Electronics, Shandong Normal University, Jinan 250014, China}

\affil[2]{Collaborative Innovation Center of Light Manipulations and Applications, Shandong Normal University, Jinan 250358, China}

\affil[3]{Department of Electrical and Computer Engineering, Dalhousie University, Halifax, Nova Scotia, B3J 2X4, Canada}

\affil[4]{Department of Physics and Atmospheric Science, Dalhousie University, Halifax, Nova Scotia, B3H 4R2, Canada}

\affil[5]{School of Physical Science and Technology, Soochow University, Suzhou 215006, China}

\maketitle
\clearpage
\begin{spacing}{1.4}
\section*{Abstract}
In the age of information explosion, the conventional optical communication protocols are rapidly reaching the limits of their capacity, as almost all available degrees of freedom (e.g., wavelength, polarization) for division multiplexing have been explored to date. Recent advances in coherent mode division multiplexing have greatly facilitated high-speed optical communications and secure, high-capacity information storage and transfer. However, coherent mode division multiplexing is quite vulnerable to even minute environmental disturbances which can cause significant information loss. Here, we propose and experimentally demonstrate a paradigm shift to incoherent mode division multiplexing for high-security optical information encryption by harnessing the degree of coherence of structured random light beams. In contrast to the conventional techniques, our approach does not require mode orthogonality to circumnavigate unwanted mode crosstalk. In addition, our protocol has, in principle, no upper bound on its capacity. Thanks to the extreme robustness of structured random light to external perturbations, we are able to achieve highly accurate information encryption and decryption in the adverse environment. The proposed protocol opens new horizons in an array of fields, such as optical imaging, optical communications, and cryptography, and it can be relevant for information processing with acoustical, seismic, matter as well as other types of waves.

\section*{Introduction}\label{sec1}

Light is the basis for our exploration of the world. Therefore, it is not surprising that the search for pathways to engineer and control the properties of light waves in space and time, from low to high dimensional settings and from classical to quantum realms, has been at the forefront of research in optical physics \cite{HechtJ}. Structured light, characterized by a tailored spatial and/or temporal distribution of the optical field and/or polarization, has emerged as a powerful concept both from the fundamental science perspective \cite{ForbesA1,WanC,ForbesA2,KondakciH} and in view of numerous promising applications \cite{OtteE,KaufmanJ,KongL}. To date, the structured light beam engineering has been largely limited to fully spatially and temporarily coherent beams \cite{ForbesA2}. The salient properties of the latter have been known to be quite sensitive to external perturbations, though. Mitigating this vulnerability has been an open challenge, requiring in-depth research. Multifarious approaches have been proposed to this end, including adaptive optics, iterative methods, deep learning models, and time-reversal techniques, among others \cite{JiN,LiM,LohaniS,VellekoopI,XuX}. Unfortunately, adaptive optics is only efficient to remedy the effects of weak perturbations in coherent mode division multiplexing protocols, while the other techniques have proven to be very time and/or resource consuming. 

At the same time, research on structured random light beams, which can be regarded as dynamic speckle patterns of light, suggests that such beams can be largely immune to noise due to the turbulence, distortions, or aberrations, originating either in the medium or at the light source \cite{HuangZ,BenderN,LiuY1,LiuY2,PonomarenkoS1,XuZ}. Therefore, any optical encryption/decryption protocol with structured random light beams, propagating in an adverse environment, requires neither pre- nor post-correction. The utilization of structured random light has already led to remarkable advances in optical image transmission through random media, super-resolution imaging, nanoparticle trapping, and optical communications \cite{BenderN,LiuY1,LiuY2,PonomarenkoS1,XuZ,PengD,ChengJ,KagalwalaK,ClarkJ,AbbeyB,HuangW,YessenovM,LumerY}.
 
Optical communications, in particular, which have greatly improved the world interconnectivity, have been experiencing a relentless pressure of ever increasing demand for channel information capacity enhancement \cite{KillingerD,KhalighiM}. To meet this demand, numerous space, time, wavelength, and polarization division multiplexing techniques have been employed to date \cite{LeiT,OuyangM}. In this context, a coherent mode division multiplexing protocol, associated with the orbital angular momentum (OAM) states of light, has lately attracted intense interest, chiefly thanks to the absence of any theoretical limit to the number of modes (OAM states) that can potentially be employed \cite{BozinovicN,WangJ}. In practice, however, such a protocol requires the use of a multitude of mutually orthogonal patterns (states) as information carriers, which is difficult to implement due to inevitable mode crosstalk. Furthermore, the existing optical communication protocols are rapidly reaching the limits of their capacity as virtually all available degrees of freedom (DoF)s of coherent orthogonal multiplexing have already been explored \cite{BozinovicN}. 

To overcome an impending gridlock in optical information processing and optical communications, we propose and experimentally realize a conceptually new division multiplexing paradigm which we refer to as incoherent mode division multiplexing. Multiplexing normalized correlation functions of the fields of a single light beam at pairs of space-time points is at the heart of our protocol. These correlation functions, known as degrees of coherence \cite{MandelL,PonomarenkoS2}, serve as generalized modes in our protocol in lieu of the orthogonal angular momentum states of light, for instance. The generalized modes are uncorrelated, thereby eliminating any mode crosstalk. In addition, the novel DoFs are known to be resilient to the environmental noise, be it the medium turbulence, or the presence of opaque obstacles to light propagation \cite{BenderN,LiuY1,LiuY2,PonomarenkoS1}. We employ the principles of optical holography to customize on demand the degree of coherence of a random light beam with the aid of a spatial light modulator (SLM). As a matter of fact, the SLM in our protocol can be replaced with a metasurface, for instance, which enables the generation of structured random light beams on demand as well \cite{LiuL}. Given its ultra-wide bandwidth and nearly universal applicability to waves of any physical nature, the incorporation of the nanophotonics technology into our protocol can allow its seamless adaptation to electromagnetic waves outside of the optical spectral range as well as to acoustical, seismic, and matter waves to mention but a few exciting possibilities \cite{RenH1,GuoX,GlybovskiS,SiderisS,AssouarB,XiaoW,LiuW}.

\section*{Results}\label{sec2}

\begin{figure}[t]
\centering
\includegraphics[width=\textwidth]{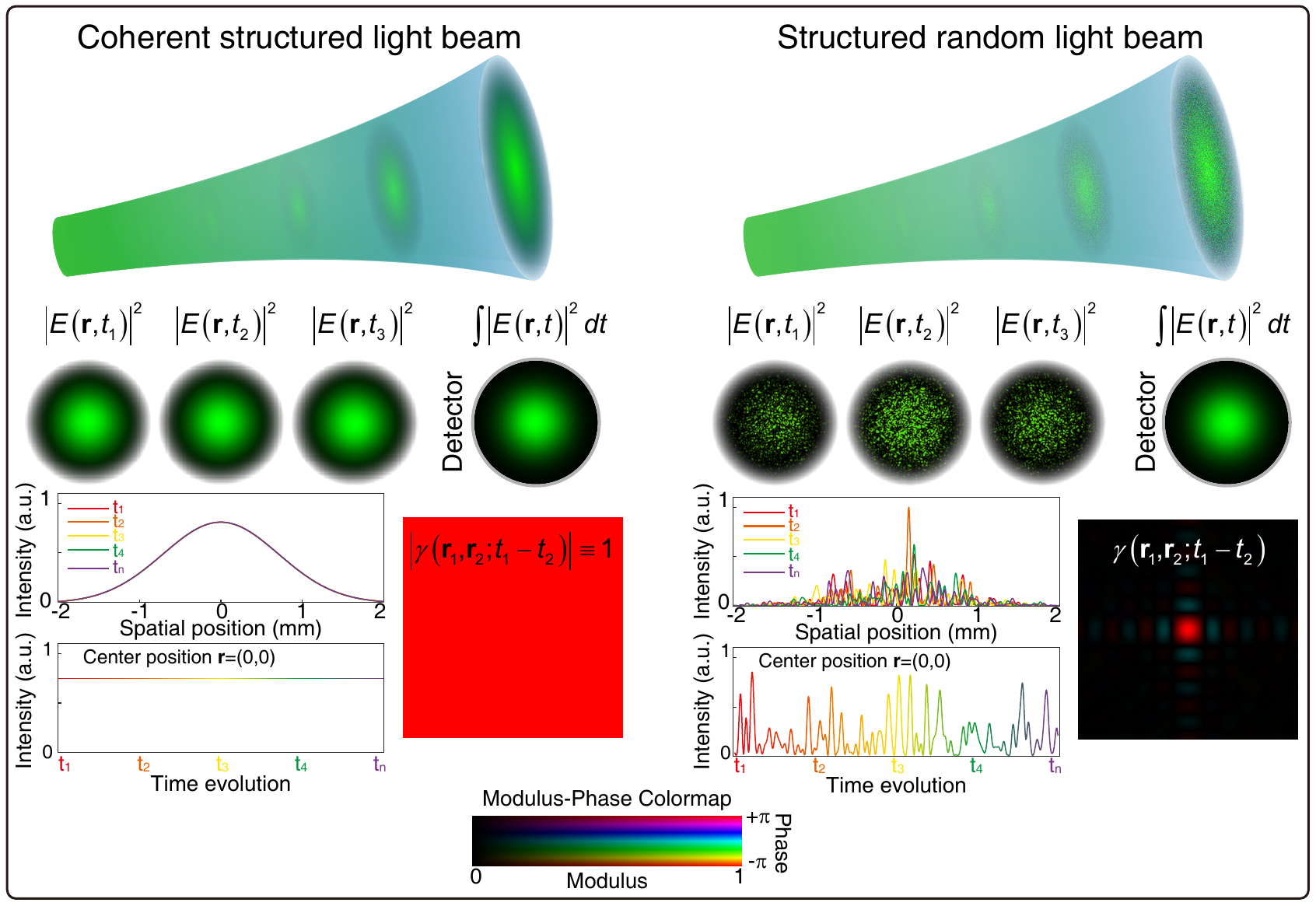}
\caption{\textbf{Comparison of coherent structured and structured random light beams.} A detecting device typically records a time-averaged intensity profile $I({\bf{r}}) \propto \int{dt}{{\lvert E( {\bf{r}},t) \lvert}^{2}}$. A coherent light beam is represented by, a stationary deterministic electric field, and unimodular degree of coherence $\lvert \gamma({\bf{r}}_1,{\bf{r}}_2;t_1-t_2)\lvert \equiv 1$. The electric field of a random light beam fluctuates in space and time, revealing a statistically granular space-time intensity distribution. The speckles are smoothed out upon time averaging, provided the integration time of the measuring device is much longer than the speckle coherence time. The degree of coherence of a structured random light beam can be readily controlled; an example is exhibited in the lower right corner, where brightness and hue quantify the modulus and phase of the degree of coherence.}\label{fig1}
\end{figure}

\bmhead{Optical coherence as additional degree of freedom} The wave properties of coherent structured light beams can be quantified by the amplitude, phase and polarization of a stationary electric field (left panel of Fig. \ref{fig1}). The structured random light beams, which can be visualized as dynamic random speckles, are described by statistically stationary electric fields which fluctuate in space and time (right panel of Fig. \ref{fig1}). Typically, a characteristic coherence time of an individual speckle is much shorter than the integration time of a detecting device. Hence, the device captures a time-averaged intensity of the beam, $I({\bf{r}}) \propto \int{dt}{{\lvert E( {\bf{r}},t) \lvert}^{2}}$, which contains no random fluctuations (right panel of Fig. \ref{fig1}). We notice that an additional DoF can be introduced to characterize coherence properties of a statistically stationary light beam, namely, the degree of coherence $\gamma({{\bf{r}}_1},{{\bf{r}}_2};{{t}_{1}}-{{t}_{2}})$ of the electric fields of the beam at a pair of space-time points $({\bf{r}}_1, t_1)$, $({\bf{r}}_2, t_2)$ (right panel of Fig. \ref{fig1}). In the fully coherent limit, $\lvert \gamma({\bf{r}}_1,{\bf{r}}_2;t_1-t_2)\lvert \equiv 1$ \cite{MandelL,PonomarenkoS2}. It follows from optical coherence theory \cite{MandelL,GoriF}, that the equal-time degree of coherence of an ensemble of bona fide random beams can be expressed as
\begin{equation}
      \gamma \left( {{\bf{r}}_{1}},{{\bf{r}}_{2}};0 \right)=\iint{p\left( {\bf{v}} \right)K\left( {\bf{{r}}_{1}},{\bf{v}} \right){{K}^{*}}\left( {{\bf{{r}}}_{2}},{\bf{v}} \right){{d}^{2}}{\bf{v}}}.
      \label{eq1}
\end{equation}
Hereafter, we restrict ourselves to the equal-time degree of coherence of an ensemble of scalar random fields to focus on spatial variables and simplify the discussion. We will refer to it as just the degree of coherence and we will drop the temporal argument of the degree of coherence for brevity. Next, $p(\bf{v})$ in Eq. (\ref{eq1}) is a (nonnegative) power spectrum density normalized to unity, $\iint{p(\bf{v})}{{d}^{2}}{\bf v}=1$ and $K(\bf{r},\bf{v})$ is an arbitrary unimodular complex function.  Further, $\bf{r}$ and $\bf{v}$ stand for radius vectors in space and reciprocal space, respectively. 

\vspace{-0.5cm}
\bmhead{Customizing structured random light beams} According to optical coherence theory \cite{MandelL}, the equal-time correlation function describing the second-order statistical properties of an ensemble of structured random light beams is defined as $\Gamma({\bf{r}}_1,{\bf{r}}_2)=\langle E^{\ast}({\bf{r}}_1){{E}}({\bf{r}}_2)\rangle$, where the electric field of any ensemble member can be specified as $E({\bf{r}})=\tau ({\bf{r}})\exp[ i\varphi ({\bf{r}})]T({\bf{r}})$. Hereafter, the angular brackets denote ensemble averaging. Further, $\tau({\bf{r}})$ and $\varphi({\bf{r}})$ are deterministic amplitude and phase, respectively. The presence of colored noise, characterized by a unimodular complex random amplitude $T({\bf{r}})$, distinguishes any structured random light beam from its coherent cousin. The degree of coherence of the ensemble is then given by $\gamma({\bf{r}}_1,{\bf{r}}_2)=\langle T^{*}({\bf{r}}_1)T({\bf{r}}_2) \rangle$. Thus, we can customize structured random light beams by designing three independent functions $\tau({\bf{r}})$, $\varphi({\bf{r}})$ and $T({\bf{r}})$ [Supplementary Figure S1]. We refer the reader to the Methods section where we elucidate our method to generate $T({\bf{r}})$.

In our experiment, we adopt a complex-amplitude modulation encoding algorithm to encode an ensemble of complex random electric fields $\left\{ E({\bf{r}}) \right\}$ into computer-generated holograms (CGHs). We refresh random matrices, composed of unimodular white noise phasors $\Im({\bf{v}})$ that are involved in generating a colored noise ensemble $\left\{ T({\bf{r}}) \right\}$, to refresh individual phase holograms. We combine all CGHs into one video. We then illuminate the CGH video, loaded onto a commercial SLM, and the first-order diffracted beams from the SLM grating constitute the sought dynamic speckles (structured random light beams). To fully characterize thus generated structured random light, we employ a perturbed Fourier intensity method to recover the degree of coherence of light. We provide further details of our experimental procedure and structured light characterization in the Methods. To demonstrate the feasibility of the method, we display experimental results for a random source example in Supplementary Figure S2. 

\begin{figure}[h]
\centering
\includegraphics[width=\textwidth]{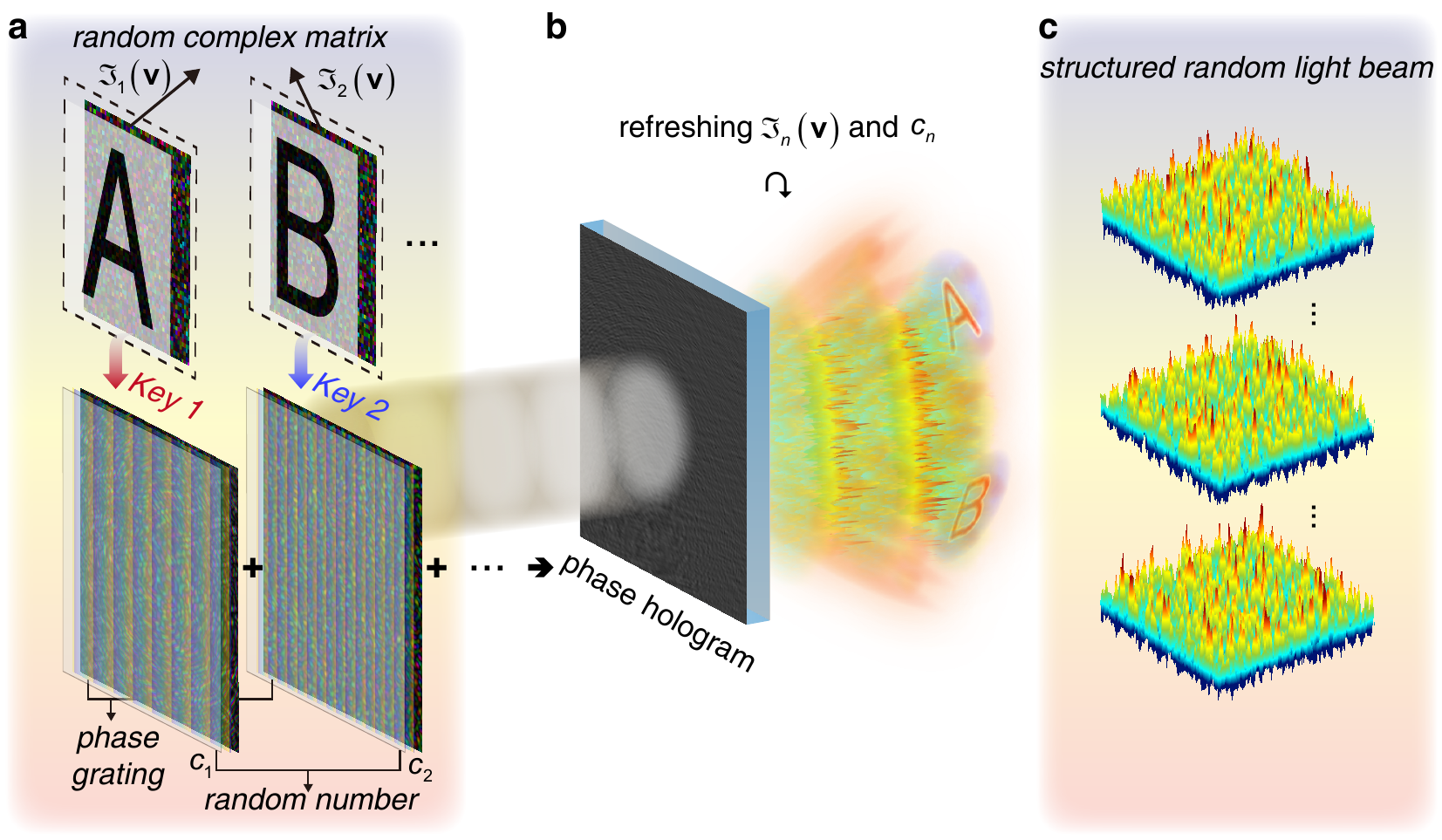}
\caption{\textbf{Hologram design for incoherent mode encoding and multiplexing.} {\textbf{a.}} and {\textbf{b.}} Schematics of a protocol for encoding multiple optical images into a structured random light beam based on Eq. (\ref{eq2}). The independent target optical images “A” and “B” are multiplied by complex random functions $\Im_{1}({\bf{v}})$ and $\Im_{2}({\bf{v}})$, and encoded into random electric fields by adopting the corresponding encryption keys (see Methods). Following a complex-amplitude modulation encoding algorithm, outlined in the Methods, each member of the random field ensemble is encoded into the individual phase grating. All phase gratings are combined into a single phase hologram by introducing a set of uncorrelated random numbers $c_n$ into each channel to eliminate mode crosstalk. {\textbf{c.}} Instantaneous speckles of structured random light beams are reproduced from the holograms by refreshing the random complex functions $\Im_{n}({\bf{v}})$ and random numbers $c_n$.}\label{fig2}
\end{figure}

\vspace{-0.5cm}
\bmhead{Incoherent mode encoding and multiplexing protocol} In the proposed protocol, the power spectrum density $p({\bf{v}})$ contains pre-encoded optical information, whereas the encoding rules (keys) are specified by $K({\bf{r}},{\bf{v}})$. Given $p({\bf{v}})$ and $K({\bf{r}},{\bf{v}})$, we can evaluate the degree of coherence of an ensemble of random beams with the aid of Eq. (\ref{eq1}). We then introduce an extended ensemble composed of subensembles  generated with different choices of $p$ and $K$. The degree of coherence of each subensemble serves as a generalized mode encapsulating a ciphertext. We can then introduce a multiplexed degree of coherence of the extended ensemble by the expression
\begin{equation}
{{\gamma }_{\mathrm{multi}}}\left({\bf{r}}_1,{\bf{r}}_2\right)={\sum\limits_{n}{\overline{c_{n}^{2}}{{\gamma }_{n}}\left( {\bf{r}}_1,{\bf{r}}_2 \right)}}/{\sum\limits_{n}{\overline{c_{n}^{2}}}}\;.
\label{eq2}
\end{equation}
Here $\{c_n\}$ is a set of real uncorrelated random numbers of unit variance representing relative weights of the generalized modes. The bar denotes averaging over the set. We note that as the number of subensembles is, in principle, unlimited, so is the number of DoFs in our protocol. Furthermore, the generalized modes are uncorrelated to ensure there is no mode crosstalk. In Fig. \ref{fig2}, we visualize our protocol for incoherent mode encoding and multiplexing. First, we encrypt optical images into each degree of coherence function $\gamma_n$ using Eq. (\ref{eq1}) with the corresponding encryption key $K_n$. Next, we generate the corresponding colored noise amplitude $T_{n}({\bf{r}})$, as is illustrated in the bottom left corner of Fig. \ref{fig2}a and explained in the Methods. An ensemble representation $T_{\mathrm{multi}}({\bf{r}})$ of the multiplexed colored noise field can be expressed as ${{T}_{\mathrm{multi}}}( {\bf{r}})={\sum\nolimits_{n}{{{c}_{n}}{{T}_{n}}(\bf{r})}}/{\sqrt{\sum\nolimits_{n}{\overline{c_{n}^{2}}}}}\;$. Assuming, for simplicity, that the deterministic amplitude $\tau({\bf{r}})$ is unity and phase $\varphi({\bf{r}})$ is constant, we can encode the resulting complex random electric field ${{E}_{\mathrm{multi}}}({\bf{r}})\propto {{T}_{\mathrm{multi}}}({\bf{r}})$ with the degree of coherence ${{\gamma }_{\mathrm{multi}}}( {{\bf{r}}_{1}},{{\bf{r}}_{2}})$, given by Eq. (\ref{eq2}), into a phase hologram via the complex-amplitude modulation encoding algorithm [see Fig. \ref{fig2}b]. We refresh random numbers $c_n$ and auxiliary random functions $\Im_n({\bf{v}})$, which describe white noise amplitudes necessary to construct colored noise amplitudes as is elaborated in the Methods, to refresh phase holograms. Repeating the protocol, we amass an ensemble of speckles encapsulating the multiplexed optical information, which is schematically shown in Fig. \ref{fig2}c. 

\begin{figure}[h!]
\centering
\includegraphics[width=\textwidth]{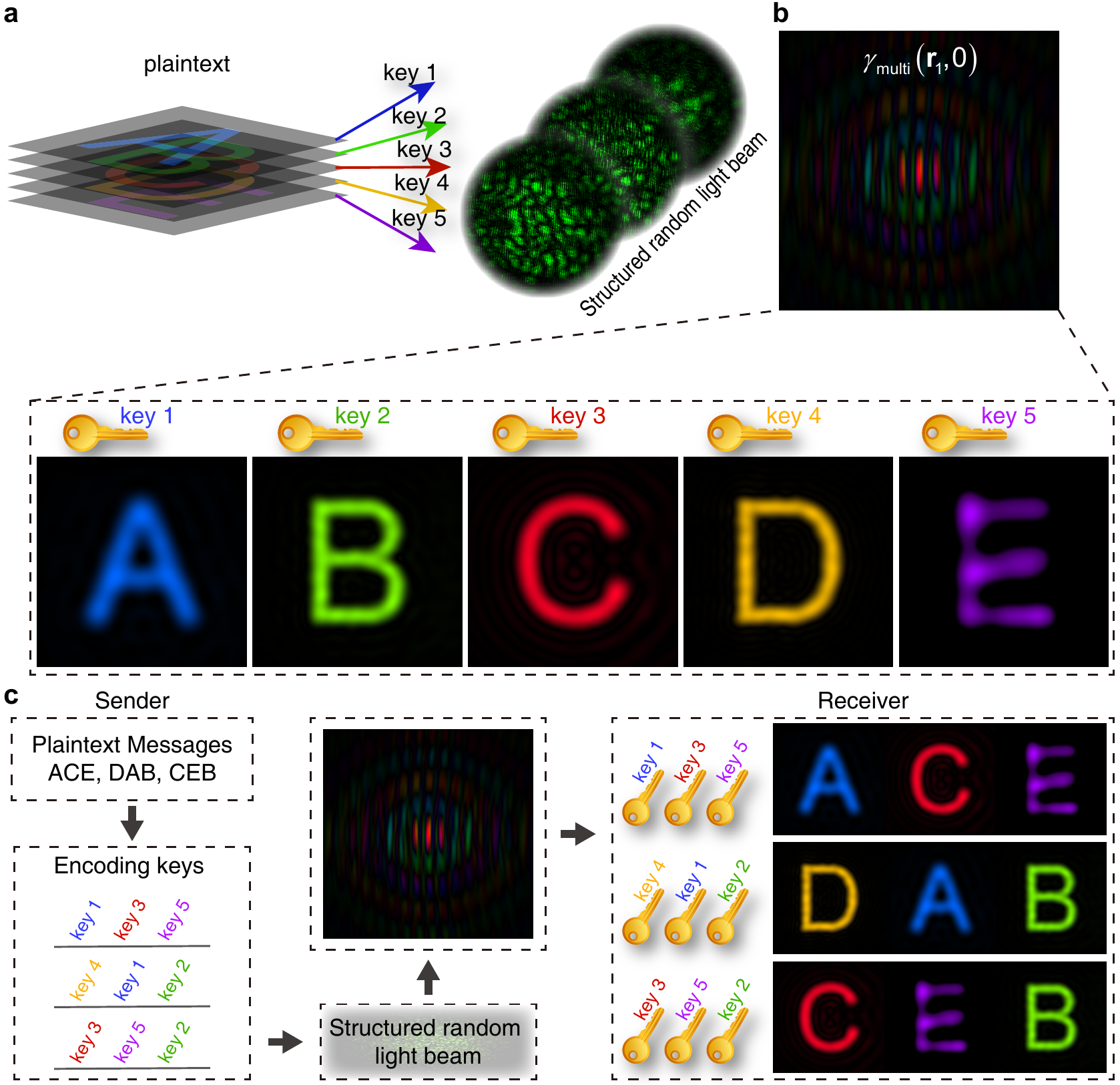}
\caption{\textbf{Proof-of-concept experimental demonstration of employing a structured random light beam to implement optical encryption and decryption.} {\textbf{a.}} The plaintexts (“A”, “B”, “C”, “D”, and “E”) are encoded into structured random light beams (some of the dynamic speckle patterns are displayed on the right). {\textbf{b.}} The multiplexed degree of coherence of the beam (the ciphertext) is assessed with the aid of the perturbed Fourier intensity method. The plaintexts are extracted from the multiplexed degree of coherence of light with respective matched encryption keys. {\textbf{c.}} A high-security and high-capacity communication protocol between sender and receiver is exhibited. A sender encodes plaintext messages into the multiplexed degree of coherence of light (the ciphertext). A receiver captures the beam and recovers the multiplexed degree of coherence. The sender’s plaintext messages are decrypted with the help of a matching decryption key sequence.}\label{fig3}
\end{figure}

In Fig. \ref{fig3}a, we display the plaintexts “A”, “B”, “C”, “D”, and “E” that are encoded into the dynamic speckles following the encryption rules described in the Supplementary Material. We provide the cumulative CGH video and generated dynamic speckles there as well (Supplementary Videos 1-1 and 1-2, respectively). Further, we list our customized encryption keys in Supplementary Table 1. We employ the perturbed Fourier intensity method to obtain the multiplexed degree of coherence of the structured random light beam ensemble. We exhibit our experimental results in the upper right corner of Fig. \ref{fig3}. We employ said encryption keys to extract the plaintexts—technical details on how to decrypt the plaintext from the measured degree of coherence are presented in Supplementary Note 1—and we show the experimental decryption results at the bottom of Fig. \ref{fig3}b. A strict one-to-one mapping of the encryption rules and decryption results ensures high security of the adopted code chart. 

Informed by our customized code chart, we propose the following high-security communication protocol between a sender (Tom) and receiver (Amy). Let us assume Tom wants to send a set of messages “ACE”, “DAB” and “CEB” to Amy. According to the code chart in Fig. \ref{fig3}b, he first defined encryption key sequences for his messages and encodes all letters into the multiplexed degree of coherence of a structured random light beam (left panel column of Fig. \ref{fig3}c). At the receiver end, Amy captures the beam and employs the perturbed Fourier intensity method to recover the multiplexed degree of coherence of the beam field. She can successfully extract optical information with the defined encryption key sequences (right panel column of Fig. \ref{fig3}c). In addition, a chronological set of encryption keys can enable the encryption and decryption of a video message via the incoherent mode multiplexing protocol as is illustrated in Supplementary Video 2. 

\begin{figure}[t]
\centering
\includegraphics[width=\textwidth]{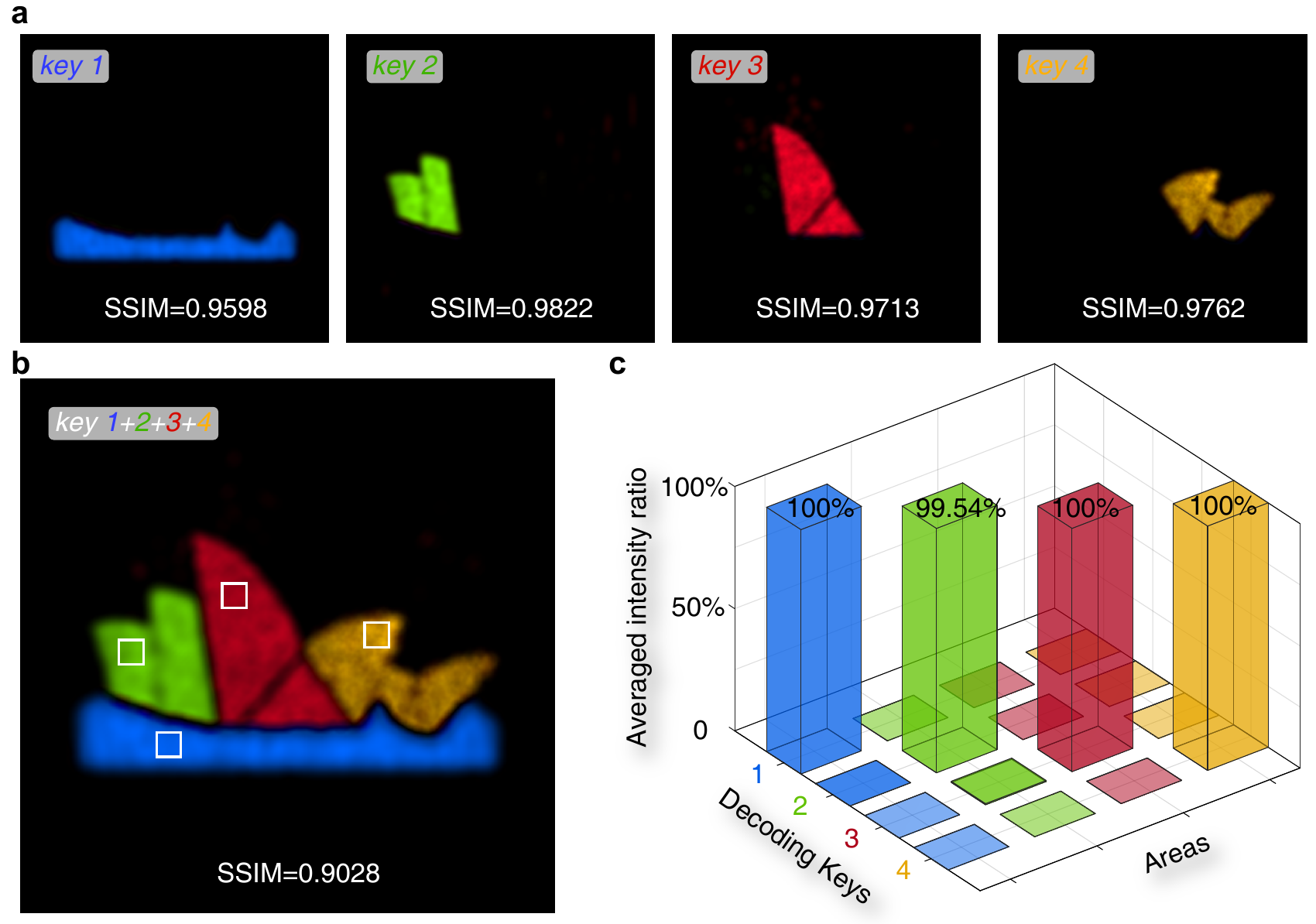}
\caption{\textbf{Incoherent mode division multiplexing for target image spatial multiplexing.} {\textbf{a.}} Experimental decryption of individual segments of the Sydney Opera House image from a structured random light beam using matched encrypted keys. {\textbf{b.}} The image of the entire Sydney Opera House is decoded as we apply all encryption keys simultaneously. {\textbf{c.}} Decoded channel crosstalk analysis of randomly selected areas [marked by the white rectangles in part {\textbf{b}}]. The SSIM index adopted to evaluate the quality of image recovery is given in each panel.}\label{fig4}
\end{figure}

Up to this point, we have demonstrated how to extract information by applying the encryption keys in a chronological sequence. By the same token, we can apply all our encryption keys at once to extract spatially separated multipath information. To illustrate this possibility, we divide a target image, the Sydney Opera House, say, into four individual segments [Supplementary Figure S3(a)]. We chose this example to demonstrate that our protocol is, at least on par with, if not superior to the coherent mode division multiplexing technique (OAM multiplexing holography) developed in \cite{FangX,RenH2,RenH3}. We encode each segment of the target image into a generalized incoherent mode (i.e., the degree of coherence of a subensemble) of an ensemble of random light beams [Supplementary Video 3] via customized encoding keys [Supplementary Table 2]. We display our experimental results for the multiplexed degree of coherence of the ensemble in Supplementary Figure S3(b). Next, we show in Fig. \ref{fig4}a four individual segments of the Sydney Opera House image which are reconstructed from Supplementary Figure S3(b) by adopting matched encryption keys. If all encryption keys, that is $key=key1+key2+key3+key4$, are applied simultaneously, the entire image of the Sydney Opera House can be faithfully reproduced (Fig. \ref{fig4}b). To evaluate the quality of the recovered images, we adopt a structural similarity index (SSIM, Supplementary Note 2) which falls in the interval [0, 1]. The greater the index, the better the image recovery. The calculated SSIM indices are given at the bottom of each panel to Fig. \ref{fig4}(a,b). The SSIM indices are seen to be high. Moreover, the image quality can be boosted even further by increasing the size of the speckle ensemble. We have also quantitatively analyzed the channel crosstalk for the image segment reconstruction among randomly selected pixels that are marked by white rectangles in Fig. \ref{fig4}b. The averaged intensity ratios of the segments, reconstructed with the corresponding encryption keys, are nearly equal to unity, and the mismatched ones are virtually equal to zero, as is illustrated in Fig. \ref{fig4}c. We can infer from Fig. \ref{fig4}c that channel crosstalk in our protocol is negligible. More importantly, our protocol makes it possible to encode and decode color images (24 bits or more). To this end, we only require three uncorrelated modes to represent three primary colors (red, green, and blue) and a single transmission of the ciphertext (i.e., multiplexed degree of coherence) between the sender and receiver [Supplementary Figure S4] regardless of the image size in terms of the number of pixels. This makes our protocol significantly faster and less resource intensive than the competition. Consider, for the sake of comparison, a recently introduced OAM multiplexing \cite{GongL,ZhaoQ}.  First, the latter calls for 24 OAM states to encode 24-bit pixels. Second, each image pixel has to be transmitted separately, making the entire image transmission very time consuming.

\begin{figure}[t]
\centering
\includegraphics[width=\textwidth]{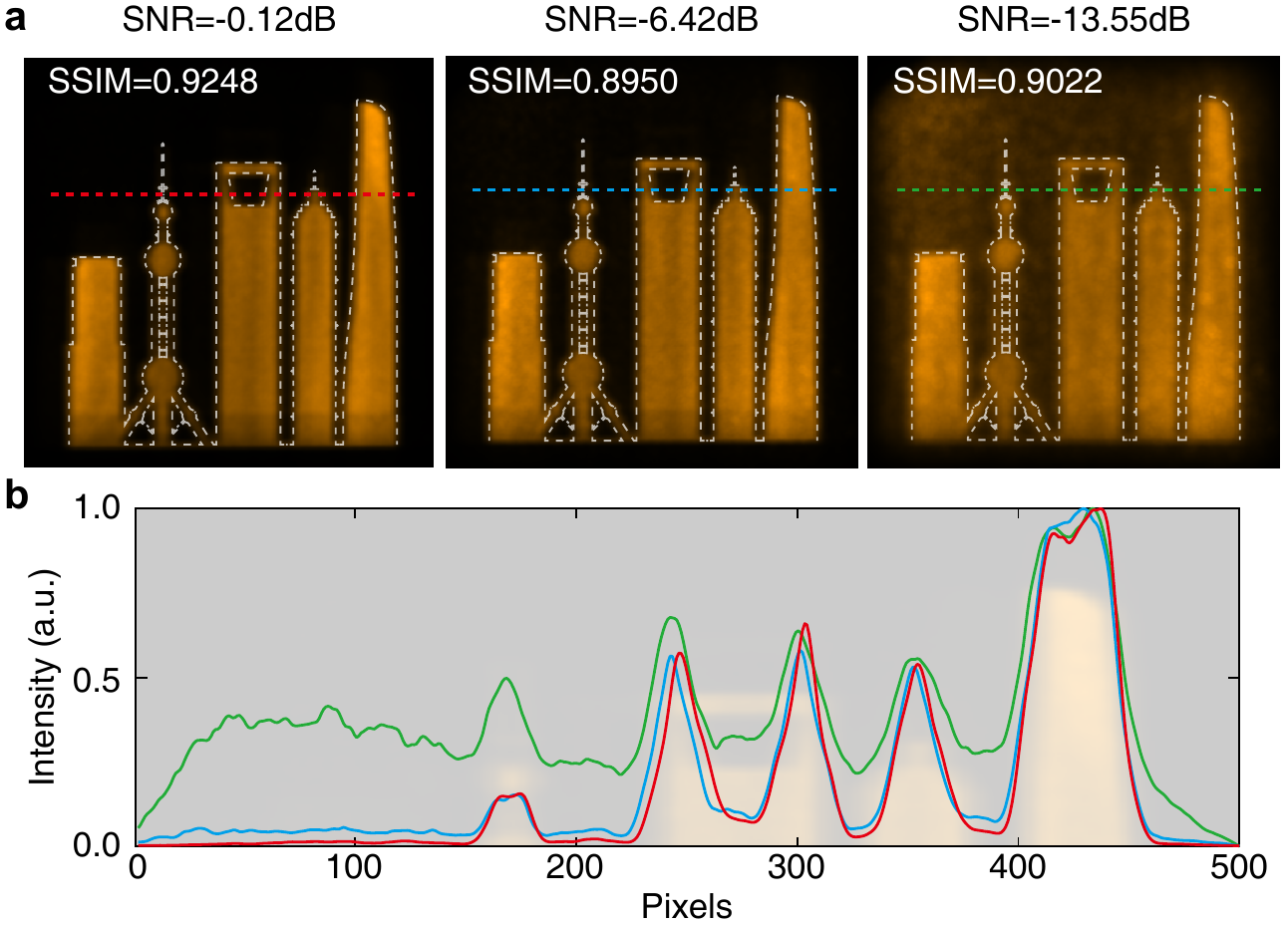}
\caption{\textbf{Robustness analysis of the decoded results from incoherent modes in a noise environment.} {\textbf{a.}} Decoded object images of the Oriental Pearl TV Tower in a white noise environment with different noise levels. {\textbf{b.}} Normalized visibility curves (cross sections) of the decoded images along the dashed lines indicated in {\textbf{a}}.}
\label{fig5}
\end{figure}

\bmhead{Protocol robustness against noise} Structured random light beams are known to be robust against disturbances caused by environmental noise, including atmospheric turbulence \cite{PonomarenkoS1,XuZ}. We now demonstrate the key advantage of our protocol over the competition—its remarkable resilience to noise. Noise is ubiquitous in nature. In optical communications, imaging and encryption, noise commonly originates from either a light source, or the refractive index fluctuations in the medium, or else, a detecting device (camera image noise, including thermal noise, {\lq\lq fixed pattern\rq\rq} noise, banding noise, etc.). We consider white noise as a generic example. To evaluate the resilience of our protocol to white noise, we adopt the following signal-to-noise ratio (SNR):
\begin{equation}
\mathrm{SNR}=10{{\log }_{10}}\left( \frac{\int{{{I}^{\mathrm{Signal}}}}{{d}^{2}}{\bf{r}}}{\int{{{I}^{\mathrm{Noise}}}}{{d}^{2}}{\bf{r}}} \right),
\label{eq3}
\end{equation}
where $I^{\mathrm{Signal}}$ and $I^{\mathrm{Noise}}$ denote signal and noise intensities of a customized structured random light source, respectively. The signal intensity corresponds to an elaborate image of the Oriental Pearl TV Tower buildings in Shanghai. The assosiated encoding key list is given in Supplementary Table 3. We capture random light beams with different SNRs in Supplementary Video 4. We also display the corresponding measured degrees of coherence affected by noise in Supplementary Figure S5. We present numerical results for the recovered image with different white noise levels in Supplementary Figure S6. In Fig. \ref{fig5}, we exhibit our recovered images (top row) and their normalized visibility curves (bottom panel) that correspond to the locations marked by the colored dashed lines in the top row. The recovered images for SNR=-0.12dB and -6.42dB (signal energy is 49.31\% and 18.57\% of the total energy, respectively) have a high resolution. Remarkably, even as SNR drops to -13.55dB (signal energy is as low as about 4.2\% of the total energy), the target images are still recovered well, though the image background appears a little hazy. We can eliminate the background noise to a certain extent by the zeroing operation. Finally, we calculate the SSIM indices of the simulated and experimental results for the recovered images and present them in each panel to the figure. The indices can reach values as high as 0.92 and 0.9 for the simulated (see Supplementary Figure S6) and experimental (see Fig. \ref{fig5}) results, respectively. It is worth noting that the SSIM index only slightly decreases from 0.9314 to 0.9077  as the fraction of signal energy relative to noise drops precipitously from 100\% to 4.2\% (Supplementary Figure S6). The experimental results are quite consistent with the simulated ones. We can then conclude that our protocol is robust even against strong noise, opening the possibility to implement incoherent mode division multiplexing for optical information encoding and transfer in complex environments, including the turbulent atmosphere in the strong fluctuation regime \cite{AndrewsL}. 
\vspace{-0.25cm}
\section*{Discussion}\label{sec12}
We have proposed and experimentally verified the concept of incoherent mode division multiplexing. In our protocol, the complex degree of coherence of an ensemble of structured random light beams plays the role of a generalized mode, or degree of freedom. In theory, the capacity of the proposed protocol has no upper bound thanks to the availability of an unlimited number of generalized modes. Our experiments have confirmed that our protocol enables the implementation of high-precision, high-security, ultrahigh information capacity optical information/image encoding which is extremely robust against noise. The proposed incoherent mode division multiplexing can be readily extended to incorporate additional DoFs for optical information encryption and transmission. For example, we can employ the wavelength or polarization division multiplexing with our incoherent mode division multiplexing to enhance the information multiplexing capacity even further. It also is worth noting that our encryption keys can be customized in arbitrary forms, which significantly broadens the applicability of the proposed incoherent mode multiplexing technique. We fully expect that we can improve the encrypted information security by increasing the complexity of encryption keys. As a matter of fact, our protocol guarantees high security even with a simple set of encryption keys because the ciphertext, represented by the degree of coherence of a random ensemble which is a second-order correlation function of light fields, cannot be directly captured by a detecting device. We have to accurately reconstruct both the magnitude and phase of the degree of coherence of light from our measurements, so that the information recovery is possible after careful mathematical processing. 

Finally, we remark that the emerging nanophotonic device (e.g., metasurface) technology with its ultra-high spatial resolution, ultra-wide bandwidth, and high integrability can facilitate structuring correlations of the amplitudes and polarization states of random waves regardless of their physical nature. In particular, Liu et al. have recently reported on the generation of structured random light beams with tailored degree of coherence with the aid of optical metasurfaces \cite{LiuL}. We then conjecture that our protocol can be extended beyond the realm of optics not only to electromagnetic waves outside the visible spectral range, but also to acoustical, seismic, and even matter waves with the help of the appropriate metasurfaces.

\backmatter

\section*{Methods} \label{methods}

\bmhead{Derivation of the colored noise function $T(\bf{r})$} The degree of coherence of the field of a structured random light beam is given by Eq. (\ref{eq1}). Applying the Van Cittert-Zernike theorem \cite{MandelL}, we can rewrite it as follows 
\begin{equation}
\gamma \left( {\bf{r}}_1,{\bf{r}}_2 \right)=\iint{\iint{{{\Gamma }_{0}}\left({\bf{v}}_1,{\bf{v}}_2 \right)K\left( {\bf{r}}_1,{\bf{v}}_1 \right){{K}^{*}}\left( {\bf{r}}_2,{\bf{v}}_2 \right){{d}^{2}}{{\bf{v}}_1}}{{d}^{2}}{{\bf{v}}_2}},
\label{eq4}
\end{equation}
where 
\begin{equation}
{{\Gamma }_{0}}\left({\bf{v}}_1,{\bf{v}}_2 \right)=\sqrt{p\left( {\bf{v}}_1 \right)}\sqrt{p\left( {\bf{v}}_2 \right)}\delta \left({\bf{v}}_1 - {\bf{v}}_2\right).
\label{eq5}
\end{equation}
Eqs. (\ref{eq4}) and (\ref{eq5}) imply that the degree of coherence of the field can be obtained by examining propagation of a completely incoherent field through an optical system. Here ${\Gamma_{0}}({\bf{v}}_1,{\bf{v}}_2)$ denotes the correlation function of the incoherent field, and $K({\bf{r}},{\bf{v}})$ stands for a response function of the optical system. Alternatively, the Dirac delta-function $\delta ({\bf{v}}_1 - {\bf{v}}_2)$ can be interpreted as a correlation function of white noise. We can then write $\delta ({\bf{v}}_1 - {\bf{v}}_2)=\langle \Im ({\bf{v}}_1){{\Im }^{*}}({\bf{v}}_2)\rangle$ where we introduced an auxiliary field $\Im({\bf{v}})$ with white noise correlations.

On substituting from Eq. (\ref{eq5}) into Eq. (\ref{eq4}), the latter is reduced to $\gamma({\bf{r}}_1,{\bf{r}}_2)=\langle T({\bf{r}}_1)\times{{T}^{*}}({\bf{r}}_2)\rangle$, where a colored noise field $T({\bf{r}})$ is expressed as 
\begin{equation}
T\left( {\bf{r}} \right)=\iint{\sqrt{p\left( {\bf{v}} \right)}\Im \left( {\bf{v}} \right)K\left( {\bf{r}},{\bf{v}} \right){{d}^{2}}{\bf{v}}}.
\label{eq6}
\end{equation}
In Eq. (\ref{eq6}), $p(\bf{v})$ represents the spectrum density of noise, which, in general, is not uniform—hence the identification of $T(\bf{r})$ with colored noise. 

\bmhead{Complex amplitude modulation encoding algorithm} To encode the complex amplitude, carrying desired information, into a phase-only SLM, we follow the procedure described in \cite{RosalesC} to design CGHs. The SLM phase is given by 
\begin{equation}
{{\Phi }_{\mathrm{SLM}}}\left( {\bf{r}} \right)={{F}_{E}}\sin \left[ \text{Arg}\left[ E\left( {\bf{r}} \right) \right]+2\pi {{f}_{x}}x \right].
\label{eq7}
\end{equation}
In Eq. (\ref{eq7}), “Arg” corresponds to the phase of $E(\bf{r})$. Next, $F_E$ is obtained by numerical inversion: ${{J}_{1}}({{F}_{E}})=\text{Abs}[E]$, where $J_1$ is a Bessel function of the first kind and first order. The phase shift $2\pi f_{x}x$ is imposed by a blazed grating, where $f_x$ denotes an inverse spatial period of the grating; $f_x$ determines the angular deviation in the $x$-direction of a light beam transmitted by the SLM. To ensure that there is a one-to-one correspondence between $F_E$ and $E(\bf{r})$, $F_E$ has to be restricted to the interval [0, 1.84]. The upper limit is attained as $J_1$ reaches its maximum. We find that the phase characterized by Eq. (\ref{eq7}) does not fall within the interval $[-\pi,\pi)$. Hence, we need to adjust the gray level of CGHs. More details are provided in Supplementary Note 3. In Fig. \ref{fig6}a, we show an example of a CGH—a screenshot of the CGH video—of the random electric field. Notably, the hologram contains the amplitude and phase modulation simultaneously. The modulation depth and phase delay control the intensity and the wavefront of the incident light. We further discuss the holographic reconstruction principle in Supplementary Note 4. 

\begin{figure}[t]
\centering
\includegraphics[width=\textwidth]{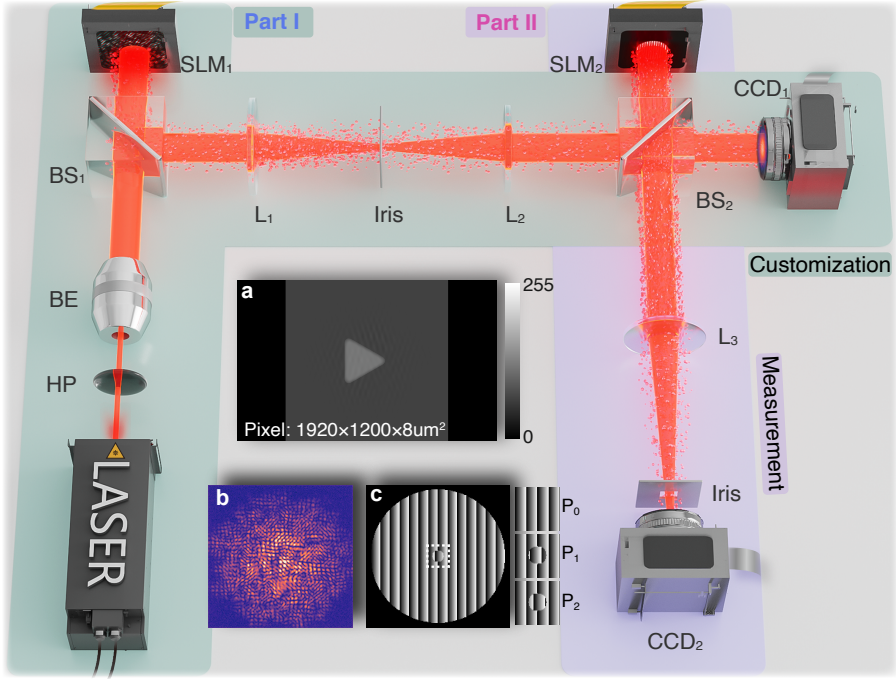}
\caption{\textbf{Experimental setup for realizing a structured random light beam [part (I) in cyan background] and its characterization [part (II) in light lavender background].} {\textbf{a.}} Screenshot of the CGHs video loaded onto the $\mathrm{SLM_{1}}$ screen; {\textbf{b.}} Example of an instantaneous intensity recorded by the $\mathrm{CCD_{1}}$ camera; {\textbf{c.}} CGHs on the $\mathrm{SLM_{2}}$ with three imparted phase patterns. HP, half-wave plate; BE, beam expander; $\mathrm{BS_{1,2}}$, beam splitters; $\mathrm{SLM_{1,2}}$, spatial light modulators; $\mathrm{L_{1,2,3}}$, thin lenses with focal lengths $f_{1,2}=15$cm and $f_{3}=10$cm; $\mathrm{CCD_{1,2}}$, charge-coupled device cameras.}
\label{fig6}
\end{figure}

\bmhead{Experimental demonstration} We sketch our experimental setup in Fig. \ref{fig6}. A linearly polarized light beam, emitted from a He-Ne Laser (THORLABS, model HNLS008L-EC) operating at a carrier wavelength of 632.8 nm is transmitted through a half-wave plate (HP) and is expanded by a beam expander (BE, with a $5\times$ magnification). The light beam is then transmitted through the first beam splitter (${\mathrm{BS_1}}$) and it illuminates a phase-only spatial light modulator 1 (${\mathrm{SLM_1}}$, Meadowlark Optics, 8-bit and $1920\times1200$ pixels with $8\,\mu m^2$). To produce high-quality optical fields, the ${\mathrm{SLM_1}}$ must be calibrated to a linear $2\pi$ phase response over the 256-gray level at wavelength 632.8nm. The prepared CGH videos are loaded on ${\mathrm{SLM_1}}$.  The reflected speckles in the first—positive or negative—diffraction order constitute our structured random light beam. We employ a 4f optical imaging system with an iris to select the desired beam. The dynamic speckles $I^{\mathrm {sp}}$ are recorded by a charge-coupled device (${\mathrm{CCD_1}}$) camera (Sony IMX252, $2048\times1536$ pixels with $3.45 \mu m^2$). We obtain a smooth intensity profile, $S\left( \mathbf{r} \right)=\int{{{I}^{\text{sp}}}\left( \mathbf{r},t \right)dt}\approx \sum\limits_{n}^{N}{{{{I}^{\text{sp}}}\left( \mathbf{r} \right)}/{N}\;}$, due to ergodicity of statistically stationary light. The equality only holds if the number of ensemble realizations $N$ is large enough, so we take $N=5000$ in our experiments. We exhibit our experimental procedure for structuring a random light beam in the module with cyan background in Fig. \ref{fig6}. 

\bmhead{Perturbed Fourier intensity method} To recover the degree of coherence of the structured random light field, we employ the following perturbed Fourier intensity method \cite{ZengJ}. We first assume that the target light beam is modulated by a phase function ${{{\mathrm O}}_{\beta}}(\bf{r})=\mathrm{M}(\bf{r})+\delta ( \bf{r}-{{\bf{r}}_{\mathrm P}} ){{\mathrm P}_{\beta}}$, where $\mathrm{M}(\bf{r})$ is a phase window to limit the area of interest and $\delta$ is a Dirac delta-function. Further, $\mathrm{P}_{\beta}(\beta=0,1,2)$ are three complex constant perturbations at $\bf r=\bf{r_{\mathrm{p}}}$, such that $\mathrm{P_0}=0$ and $\mathrm{P_1}=\mathrm{P_2^*}$. The (spectral) intensity of the modulated light beam in the Fourier domain is given by
\begin{equation}
{{\tilde{S_{\beta }}}}\left( \boldsymbol{\kappa}  \right)=\iint{\Gamma \left({\bf{r}}_1,{\bf{r}}_2 \right){{{\mathrm O}}_{\beta }}\left( {\bf{r}}_1 \right){\mathrm O}_{\beta }^{*}\left( {\bf{r}}_2 \right)\exp \left[ -i2\pi \left( {\bf{r}}_1 - {\bf{r}}_2 \right)\cdot \boldsymbol{\kappa}  \right]{{d}^{2}}{{\bf{r}}_1}{{d}^{2}}{{\bf{r}}_2}}.
\label{eq8}
\end{equation}
We obtain three spectral densities ${\tilde{S_\beta}}$ for $\mathrm{P}_{\beta}(\beta=0,1,2)$, respectively. After mathematical processing, the complex correlation function at a pair of points one of which is a reference point ${\bf{r}}_2={\bf{r}}_{\mathrm{p}}$ can be expressed as \cite{ZengJ} 
\begin{equation}
\Gamma \left( {\bf{r}}_1,{\bf{r}}_{\mathrm{p}} \right)=\frac{{{\mathrm P}_{2}}{{\mathcal{F}}^{-1}}\left[ {{{\tilde{S_{1}}}}}\left( \boldsymbol{\kappa}  \right)-{{{\tilde{S_{0}}}}}\left( \boldsymbol{\kappa}  \right) \right]-{{\mathrm P}_{1}}{{\mathcal{F}}^{-1}}\left[ {{{\tilde{S_{2}}}}}\left( \boldsymbol{\kappa}  \right)-{{{\tilde{S_{0}}}}}\left( \boldsymbol{\kappa}  \right) \right]}{{{\mathrm P}_{1}}^{*}{{\mathrm P}_{2}}-{{\mathrm P}_{2}}^{*}{{\mathrm P}_{1}}},
\label{eq9}
\end{equation}
where ${{\mathcal{F}}^{-1}}$ denotes an inverse Fourier transform. The degree of coherence is obtained by normalizing $\Gamma ({\bf{r}}_1,{\bf{r}}_{\mathrm p})$.

We exhibit the corresponding experimental setup in Fig. \ref{fig6} part (II) in the module with light lavender background. The structured random light beam reflected by $\mathrm {BS_2}$ is imaged onto the screen of the second phase-only SLM ($\mathrm {SLM_2}$, Meadowlark Optics, $1920\times1200$ pixels with $\mathrm {8\mu m^2}$) by the 4f optical imaging system. We encode the phase functions ${{{\mathrm O}}_{\beta}}(\bf{r})$ into the holograms with the values 1, $\exp(2\pi/3)$ and $\exp(-2\pi/3)$ of $\mathrm {P_0}$, $\mathrm {P_1}$ and $\mathrm {P_2}$, respectively [Fig. \ref{fig6}c]. The beam reflected by the $\mathrm {SLM_2}$, which is loaded with said holograms, is focused by the lens $\mathrm {L_3}$ and an iris is used to select the first positive diffraction order (the desired perturbed field). The $\mathrm {CCD_2}$ (Sony IMX252, $2048\times1536$ pixels with $\mathrm {3.45\mu m^2}$) camera is located in the rear focal plane of the lens, whereas the $\mathrm {SLM_2}$ is situated in its front focal plane. The intensity of light captured by the $\mathrm {CCD_2}$ is given by Eq. (\ref{eq8}). Finally, we can obtain the degree of coherence of the structured random light field with the aid of Eq. (\ref{eq9}). 

\end{spacing}

\section*{Data availability}
\noindent{The data supporting the findings of this study are available from the corresponding authors upon reasonable request.}

\section*{Code availability}
\noindent{The code producing the figures is available from the corresponding authors upon reasonable request.}

\section*{Acknowledgments}
This work was supported by the National Key Research and Development Program of China (2022YFA1404800, 2019YFA0705000), National Natural Science Foundation of China (12004220, 12192254, 11974218), Regional Science and Technology Development Project of the Central Government (No. YDZX20203700001766), China Postdoctoral Science Foundation (2022T150392, 2019M662424), the Natural Sciences and Engineering Research Council of Canada (RGPIN-2018-05497).

\section*{Author contributions}
Y.C. and C.L. conceived the idea and initiated the research project. X.L and S.A.P developed the theoretical framework. X.L. designed and implemented the experiment. F.W. analyzed the data. All authors discussed the results and contributed to the text of the manuscript.

\section*{Competing interests}
The authors declare no competing interests.

\section*{Additional information}
\bmhead{Supplementary information} The online version contains supplementary material available at \dots

\end{document}